\documentclass[preprint,12pt]{elsarticle}
\usepackage[margin=20mm]{geometry}
\biboptions{sort&compress}
\usepackage[T1]{fontenc}
\usepackage{ragged2e}
\usepackage{amsmath,amssymb,amsfonts,mathtools,bm}
\usepackage{amsthm}
\usepackage{xfrac}
\usepackage{mathrsfs}
\usepackage{dsfont}
\usepackage{anyfontsize}
\usepackage{graphicx}
\usepackage{xcolor}
    \definecolor{notecolor}{HTML}{300d66}
    \definecolor{tmpcolor}{HTML}{E0007B}
\usepackage[colorlinks=true]{hyperref}
\usepackage{comment}
\usepackage{refcount}
\newtheorem{theorem}{Theorem}
\newtheorem{lemma}[theorem]{Lemma}

\newtheorem{corollary}[theorem]{Corollary}
\theoremstyle{definition}

\theoremstyle{remark}

\newcommand{\appref}[1]{\hyperref[#1]{Appendix~\ref*{#1}}}
\newcommand{\secref}[1]{\hyperref[#1]{Section~\ref*{#1}}}
\newcommand{\subsecref}[1]{\hyperref[#1]{Section~\ref*{#1}}}

\newcommand{\pf}{\operatorname{pf}\mkern-1.7mu}

\newcommand{\mi}{\mathrm{i}}

\newcommand{\bdot}{\bm{\cdot}}

\newcommand{\Fm}{\widehat{\mathsf{F}}}

\newcommand{\Hir}{{\mathcal{D}}}

\newcommand{\ba}{\begin{eqnarray}}
\newcommand{\ea}{\end{eqnarray}}

\begin{document}
\hypersetup{
    linkcolor=blue,
    citecolor=blue,
    urlcolor=blue
}
\begin{frontmatter}

\title{Pfaffian solutions to the Ernst equation}
\author[1]{A. R. Melikyan}
\ead{amelik@gmail.com}

\author[1]{A. Pinzul}
\ead{aleksandr.pinzul@gmail.com}
\affiliation[1]{
	organization={Instituto de Física, Centro Internacional de Física, \\ Universidade de Brasília},
	city={Brasília},
	state={DF},
	postcode={70910-900},
	country={Brazil}
}

\begin{abstract}
We obtain a Pfaffian representation of the Yamazaki-Hori solutions to the Ernst equation. The main consequence of this result is the proof of Nakamura's conjecture relating the Ernst equation to the 2D Toda model, presented in \cite{MP2026nc}.
\end{abstract}

\end{frontmatter}
\tableofcontents

\section{Introduction}
\label{intro}
The solutions of the Ernst equation, which arises in the description of stationary axisymmetric gravitational fields, can be used to recover the metric functions (see \cite{Islam:2009no} for a review). The Yamazaki-Hori solutions are exact rational solutions parametrized by the integer (deformation) parameter \(n\), and are given as the ratio of two complex polynomials \(g_{n}=g_{n}(x,y,p,q)\) and \(f_{n}=f_{n}(x,y,p,q)\):
\begin{align}
    \xi_{n} = \frac{g_{n}}{f_{n}}. \label{intro:yh_solution}
\end{align}
Here, \(x,y\) are the prolate spheroidal coordinates, and \(p,q\) are two parameters related on the physical subspace by the relation \(p^{2}+q^{2}=1\).\footnote{Note that throughout the paper we keep, as explained below, the parameters \(p\) and \(q\) unconstrained.} This parametrization is fixed by requiring that the solution \(\xi_{1}\) corresponding to \(n=1\) reproduces the Kerr solution \cite{Islam:2009no}.
The Yamazaki-Hori solutions \eqref{intro:yh_solution} extend the solutions found earlier by Tomimatsu and Sato for \(n=1,2,3,4\) (for a general overview and a relation to other known solutions see \cite{Tomimatsu:1981fk, Vein:2013kv, Fukuyama:1995cm, Fukuyama:2010gu, Melikyan:2026ux}).
In \cite{Melikyan:2026ux}, the algebraic structure of the Yamazaki-Hori solution was investigated, and it was shown that the polynomials \(f_n,g_n\in\mathcal{R}:=\mathbb{C}[p,q,x,y]\) share a greatest common divisor \(d_n:=\gcd_{\mathcal{R}}(f_n,g_n)\). More precisely,
\begin{align}
f_n=d_n\widehat{f_n}=\mathcal{K}(n)d_n\overline{d_n},\quad
g_n=d_n\widehat{g_n},
\label{intro:main_factorization}
\end{align}
where \(\mathcal{K}(n)\) is a constant which depends only on \(n\), and the reduced polynomials \((\widehat{f_n},\widehat{g_n})\) form a pair of coprime polynomials. The proof of \eqref{intro:main_factorization} given in \cite{Melikyan:2026ux} was not, however, constructive and only the existence and the degrees of the polynomial \(d_{n}\) in \(p,q,x,y\) were established. Nevertheless, it shows that the solutions to the Ernst equation reduce to the ratio between coprime polynomials: \(\xi_n = {\widehat{g}_{n}}/{\widehat{f}_{n}}\). This reduction is essential for Nakamura's conjecture \cite{Nakamura:1993ig}, which relates the Ernst solutions to the \(\tau\)-functions of the 2D Toda model. To establish this relation explicitly, one must first determine the common factor \(d_n\) and the coprime polynomials \((\widehat{f_n},\widehat{g_n})\).

The purpose of this paper is to provide a constructive proof of the factorization \eqref{intro:main_factorization} and obtain a closed form for \(d_n\). More precisely, we construct two polynomials \(P_n\) and \(Q_n\), given explicitly by Pfaffians of antisymmetric matrices, and show that \(d_n\sim\overline{P_n}\).\footnote{Here \(\sim\) denotes equality up to a nonzero constant.} We also show that the reduced polynomials \(\widehat{f_n}\) and \(\widehat{g_n}\) are proportional to \(P_n\) and \(Q_n\), respectively. The Pfaffian structure of the Tomimatsu-Sato solutions was conjectured by Nakamura and Ohta in \cite{Nakamura:1991sv}, based on computational verification for low values of \(n\), but a general analytical proof was not obtained.

As a consequence of our analysis, we obtain the following representation of the Yamazaki-Hori solution \eqref{intro:yh_solution}:
\begin{align}
\xi_n=(2n-1)
\begin{cases}
\dfrac{Q_n}{P_n}, & n-1\text{ even},\\[8pt]
2\dfrac{Q_n}{P_n}, & n-1\text{ odd}.
\end{cases}
\label{intro:ap_solution}
\end{align}
The explicit Pfaffian representations of \(P_n\) and \(Q_n\) are derived and analyzed in Sections \ref{sec:pf},\ref{sec:pfg} and \ref{pfg:polynomial_gcd}.

The main consequence of the Pfaffian representation obtained here is the proof of Nakamura's conjecture given in \cite{MP2026nc}, where we establish the following relations between the Toda \(\tau\)-functions
\footnote{The 2D Toda equation in prolate spheroidal coordinates has the form \cite{Nakamura:1993ig}:
\(
\Fm\, (\tau_{n} \bdot \tau_{n}) = 2 \tau_{n+1}\tau_{n-1},
\)
where
\[
\Fm := \left[(x^{2} - 1)^{2} \Hir_{x}^{2} +2x(x^{2} - 1) \widehat{\partial}_{x}\right] - \left[(y^{2} - 1)^{2} \Hir_{y}^{2} + 2y(y^{2} - 1) \widehat{\partial}_{y}\right],
\]
where \(\Hir_x\) and \(\Hir_y\) denote Hirota bilinear derivatives \cite{Hirota:2009fa} whose explicit form is not relevant for the present paper.
} and the Pfaffians introduced here:
\begin{align}
\tau_{n-1}(\Phi')\sim P_n ,\ \
\tau_n(\Phi)\sim Q_n .
\label{intro:naka_pf_g}
\end{align}
Here the seed functions are:
\begin{align}
\Phi(x,y)=-px+iqy,\ \
\Phi'(x,y)=-2px(x^2-1)-2iqy(y^2-1),\label{Phi}
\end{align}
and the relations hold on the physical subspace \(p^2+q^2=1\). The proof is based on an explicit transformation between the polynomial sequences entering the Toda determinants and the Pfaffians, together with an extension of the generalized Cusick identity \cite{Vein:2013kv}.
Recent progress and a more detailed discussion of Nakamura's conjecture can also be found in \cite{Fukuyama:2026qe}.

\section{The Yamazaki–Hori solution}
\label{mm}
The original Yamazaki-Hori solutions of the Ernst equation, given in \cite{Yamazaki:1977xd, Hori:1978ie}, have a rather complex form. A simpler, but equivalent representation was given in \cite{Vein:1983if, Melikyan:2025ej}. We briefly review the necessary formulas following the notation used in \cite{Melikyan:2025ej}.
We define the following \((n+1)\times(n+1)\) matrix, depending on an arbitrary function \(h=h(x,y,p,q)\):
\begin{equation}
\label{mm:hsol_B_matrix}
\mathsf{B}_n(z;h)=
\begin{pmatrix}
\zeta_0 & \zeta_1 & \cdots & \zeta_{n-2} & \zeta_{n-1} & \mathsf{b}(0,z;h)\\
\zeta_1 & \zeta_2 & \cdots & \zeta_{n-1} & \zeta_{n}   & \mathsf{b}(1,z;h)\\
\vdots  & \vdots  & \ddots& \vdots      & \vdots      & \vdots\\
\zeta_{n-1} & \zeta_{n} & \cdots & \zeta_{2n-3} & \zeta_{2n-2} & \mathsf{b}(n-1,z;h)\\[1ex]
1 & \mathsf{a}(h) & \cdots & \mathsf{a}^{n-2}(h) & \mathsf{a}^{n-1}(h) & 0
\end{pmatrix},
\end{equation}
where \(\mathsf{a}(h):=1-h\), and the border column \(\mathsf{b}(\ell,z;h)\) can be written in terms of Gauss' hypergeometric function, or, equivalently, in the following integral form:
\begin{equation}\label{mm:hsol:b_border_hyper}
\mathsf{b}(\ell,z;h):=(-h)^{\ell}\,
{}_2F_1 \Big(\frac{1}{2},\, -\ell;\,\tfrac{3}{2};\,\tfrac{z^2}{h}\Big) = \frac{1}{z} \int_{0}^{z} \left(u^2 -h \right)^{\ell}\,du.
\end{equation}
The main block of \eqref{mm:hsol_B_matrix} is a Hankel matrix \(\mathcal{H}\) with entries \(\zeta_{m}=\zeta_{i+j-2},\, 1 \le i,j \le n\).
Defining \(\Delta:=p^{2}+q^{2}-1\), their explicit form is:
\begin{align}
\zeta_{m}(x,y,p,q;h)
=\frac{1}{m+1}\Big[
&p^{2}(x^{2}-h)^{m+1}
+q^{2}(y^{2}-h)^{m+1} -(1-h)^{m+1} -\Delta(-h)^{m+1}
\Big].
\label{mm:moments_h}
\end{align}
On the physical subspace \(p^{2}+q^{2}=1\), we have \(\Delta=0\), and \eqref{mm:moments_h} reduces to the original \(h\)-dependent moments derived in \cite {Melikyan:2025ej}.\footnote{The additional term arises when applying the difference transformation to the moments corresponding to Vein's representation, without imposing the constraint \(p^{2}+q^{2}=1\).} We keep this term as it allows us to treat \(p\) and \(q\) as independent variables, which is essential for identifying \(d_n\) as the greatest common divisor of \(f_n\) and \(g_n\) in the polynomial ring \(\mathbb{C}[p,q,x,y]\).

With these notations, the solution to the Ernst equation can be written in the form:
\begin{equation}
\label{mm:hsol:am_solution}
\xi_n=\frac{g_{n}}{f_{n}},
\end{equation}
where the polynomials \(g_{n}=g_{n}(x,y,p,q)\) and \(f_{n}=f_{n}(x,y,p,q)\) are defined as:
\begin{align}
	g_{n}&=p\,x\,\det \mathsf{B}_{n}(x;h)-\mi\,q\,y\,\det \mathsf{B}_{n}(y;h), \label{mm:g_polynomial_def}\\
	f_{n}&=\det \mathsf{B}_{n}(1;h),\label{mm:f_polynomial_def}
\end{align}
with the parameters \(p\) and \(q\) related by the constraint \(p^{2}+q^{2}=1\). The solution contains an arbitrary function \(h(x,y,p,q)\), which can be used to simplify the expressions. In particular, the choice \(h=1\) reproduces the original Yamazaki-Hori representation \cite{Yamazaki:1977xd,Hori:1978ie}, while \(h =0\) leads to the equivalent representation derived by Vein \cite{Vein:1983if}.\footnote{Although the factorization of \(f_n\) and \(g_n\) was established in \cite{Melikyan:2026ux} using Vein's representation (\(h=0\)), it remains valid for arbitrary \(h\), since the bordered determinant \(\det\mathsf{B}_n(z;h)\) is independent of \(h\).}

\subsection{The moment matrix}

It is natural to express the moments \eqref{mm:moments_h} of the Hankel matrix \(\mathcal{H}\) in terms of a weight function \(w\), following the standard approach in the theory of orthogonal polynomials:

\begin{align}
\zeta_{m}(x,y,p,q;h)
=p^{2}\int_{-h}^{\infty}u^{m}\,w(u;h)\,du,
\label{mm:moments_int}
\end{align}
where \(w(u;h)\) is the weight function associated with the moments \eqref{mm:moments_h}:
\begin{align}
w(u;h) = \theta(x^{2}-h-u)
+s^{2}\theta(y^{2}-h-u)
-\frac{1}{p^{2}}\theta(1-h-u).
\label{mm:weight_function}
\end{align}


We have also defined the parameter \(s = q/p\) which represents deviations from the integrable Weyl limit \(s = 0\), with \(p=1\) and \(q=0\). In this limit, the closed forms of the polynomials \(f_{n}\) and \(g_{n}\) are well-known \cite{Vein:2013kv,Fukuyama:1995cm,Fukuyama:2011ma}, and the corresponding solutions \(\xi_{n}\) reduce to Weyl's static gravitational solutions with deformation parameter \(n\). In this paper we will consider the general case of \(s \neq 0\).

For the moment, we keep the function \(h\) arbitrary and analyze the polynomial in the denominator of the solution \eqref{mm:hsol:am_solution}, i.e., \(f_{n}={\det \mathsf{B}_{n}(1;h)}\). First, we remove the last row of \eqref{mm:hsol_B_matrix} by performing the column operations:
\begin{align}
	C_{k} \to C_{k} - \mathsf{a}(h) C_{k-1} \label{mm:column_op_1}
\end{align}
successively for \(k = n, n-1, \dots, 2\). This operation transforms the last row to \((1, 0, \dots, 0)\), and expansion along the latter then gives:
\begin{equation}
\label{mm:f_n_h_column_op}
f_{n}= (-1)^{n} \det
\begin{pmatrix}
\widetilde{\zeta}_0 & \widetilde{\zeta}_1 & \cdots & \widetilde{\zeta}_{n-2}  & \mathsf{b}^{(f)}_{0}(h)\\
\widetilde{\zeta}_1 & \widetilde{\zeta}_2 & \cdots & \widetilde{\zeta}_{n-1}  & \mathsf{b}^{(f)}_{1}(h)\\
\vdots  & \vdots  & \ddots& \vdots    & \vdots\\
\widetilde{\zeta}_{n-1} & \widetilde{\zeta}_{n} & \cdots & \widetilde{\zeta}_{2n-3} & \mathsf{b}^{(f)}_{n-1}(h)
\end{pmatrix},
\end{equation}
where we have denoted (cf. \eqref{mm:hsol:b_border_hyper})
\begin{align}\label{mm:hsol:Fhyper}
&\mathsf{b}^{(f)}_{\ell}(h):=(-h)^{\ell}\,
{}_2F_1 \Big(\frac{1}{2},\, -\ell;\,\tfrac{3}{2};\,\tfrac{1}{h}\Big) =  \int_{0}^{1} \left(u^2 -h \right)^{\ell}\,du,\\
&\widetilde{\zeta}_{k}=\zeta_{k+1} -\mathsf{a}(h) \zeta_{k}
\end{align}
and the new moments in \eqref{mm:f_n_h_column_op}  \(\widetilde{\zeta}_{k}\) can be represented by a modified weight function \(\widetilde{w}(u;h) = [u-\mathsf{a}(h)] w(u;h)\):\footnote{Note that the modified weight function \(\widetilde{w}(u;h)\) corresponds to the Christoffel transform of the weight function \(w(u;h)\) (see Section 2.7 of \cite{Ismail-2025oj} for a detailed exposition).}
\begin{align}
\widetilde{\zeta}_{k}
=p^{2}\int_{-h}^{\infty}
u^{k}\,\widetilde{w}(u;h)\,du.
\label{mm:moments_tilde_zeta}
\end{align}
The weight function \(w(u;h)\) in \eqref{mm:weight_function} is not continuous and, for a suitable choice of the function \(h\), is exactly of the type considered in \cite{Chen:2005au, Chen:2010rl, Min:2020ld}, where it was shown that such discontinuous weight functions and the resulting orthogonal polynomials lead to the Painlev\'{e} VI equation. The consequences of this connection will be explored in \cite{Medeiros-op}.

In what follows, we set \(h=0\), which, as shown in \cite{Melikyan:2026ux}, recovers Vein's solution \cite{Vein:1983if}. The main advantage of this choice is the simpler form of the border elements \eqref{mm:hsol:Fhyper}:
\begin{align}
\mathsf{b}^{(f)}_{\ell}(0)=\lim_{h\to0}\mathsf{b}^{(f)}_{\ell}(h)=\frac{1}{2\ell+1}. \label{mm:fn_border_h_0}
\end{align}
In this case, \(f_{n}\) in \eqref{mm:f_n_h_column_op} takes the form:
\begin{equation}
\label{mm:f_n_h_zero}
f_{n}= (-1)^{n} \left(p^{2}\right)^{n-1} \det
\begin{pmatrix}
\mu_0 & \mu_1 & \cdots & \mu_{n-1} \\
\mu_1 & \mu_2 & \cdots & \mu_{n}  \\
\vdots  & \vdots  & \ddots& \vdots \\
\mu_{n-2} & \mu_{n-1} & \cdots & \mu_{2n-3} \\
1 & 1/3 & \cdots & 1/(2n-1)
\end{pmatrix},
\end{equation}
where the \(\mu_{k}\) moments are written in terms of the parameter \(s=q/p\) as follows:
\begin{align}
\mu_{k} = \int_{1}^{x^{2}}u^{k}(u-1)\,du + s^{2}\int_{1}^{y^{2}}u^{k}(u-1)\,du + \frac{\Delta}{p^{2}}\int_{0}^{1}u^{k}(u-1)\,du.
\end{align}
As before, the last term vanishes on the physical subspace \(p^{2}+q^{2}=1\), but must be kept when \(p\) and \(q\) are treated as independent variables.
Due to the simple form of the border elements in \eqref{mm:fn_border_h_0}, we perform an additional column operation to reduce the dimension of the matrix in \eqref{mm:f_n_h_zero}. Namely, the last row of \eqref{mm:f_n_h_zero} can be eliminated by the column operations:
\begin{align}
	C_{j} \to C_{j} - \frac{2j+1}{2j-1} C_{j+1}\label{mm:column_op_2},
\end{align}
successively for \(j = 1,2, \dots, n-1\) and expanding along the resulting last row. The result is:
\begin{align}
f_{n} = \frac{(-1)^{n}}{2n-1} (p^{2})^{n-1} \det {K},\label{mm:f_n_reduced}
\end{align}
where we have defined the following \( (n-1) \times (n-1) \) matrix \(K\):
\begin{align}
K_{ij}:=\left[\mu_{i+j-2} -\lambda_{j} \mu_{i+j-1} \right],\quad
\lambda_{j}:= \frac{2j+1}{2j-1}, \quad 1 \leq i,j \leq n-1.
\end{align}
Thus, by a series of elementary column operations we have reduced the original \( (n+1) \times (n+1)\) matrix \(\mathsf{B}_{n}(1;h=0)\) \eqref{mm:hsol_B_matrix}, corresponding to \(f_{n}\), to an \( (n-1) \times (n-1)\) matrix \(K\) in \eqref{mm:f_n_reduced}, which has the explicit form:
\begin{equation}
\label{mm:f_n_K_matrix}
K=
\begin{pmatrix}
    \widehat{\mu}_{0}^{(1)} & \widehat{\mu}_1^{(2)} & \cdots & \widehat{\mu}_{n-2}^{(n-1)} \\
    \widehat{\mu}_{1}^{(1)} & \widehat{\mu}_2^{(2)} & \cdots & \widehat{\mu}_{n-1}^{(n-1)}  \\
\vdots  & \vdots  & \ddots& \vdots \\
\widehat{\mu}_{n-2}^{(1)} & \widehat{\mu}_{n-1}^{(2)} & \cdots & \widehat{\mu}_{2n-4}^{(n-1)}
\end{pmatrix},
\end{equation}
Here we have denoted the elements of the matrix \(K\) by \(\widehat{\mu}_{k}^{(j)}\):
\begin{align}
	\widehat{\mu}_{k}^{(j)}=\mu_{k} - \left(\lambda_{j}\right) \mu_{k+1},\quad k=i+j-2.\label{mm:tilde_mu}
\end{align}
Note that, after the last reduction, the matrix \(K\) is no longer a Hankel matrix because of the factor \(\lambda_j\) appearing in \eqref{mm:tilde_mu}. To clearly see its structure, we separate the \(i\) and \(j\) indices, and write the elements \(\widehat{\mu}_{k}^{(j)}\) in the following explicit form:
\begin{align}
\widehat{\mu}_{k}^{(j)}
={}&\int_{1}^{x^{2}}\left\{u^{i-1}(u-1)\right\}\left\{u^{j-1}(1-\lambda_j u)\right\}\,du +s^{2}\int_{1}^{y^{2}}\left\{u^{i-1}(u-1)\right\}\left\{u^{j-1}(1-\lambda_j u)\right\}\,du \nonumber\\
&+\frac{\Delta}{p^{2}}\int_{0}^{1}\left\{u^{i-1}(u-1)\right\}\left\{u^{j-1}(1-\lambda_j u)\right\}\,du.
\end{align}
This can be more conveniently written as:
\begin{align}
K_{ij} = \int_{1}^{x^{2}}F^{(i)}G^{(j)}\,du + s^{2}\int_{1}^{y^{2}}F^{(i)}G^{(j)}\,du + \frac{\Delta}{p^{2}}\int_{0}^{1}F^{(i)}G^{(j)}\,du.
\label{mm:tilde_mu_i_j}
\end{align}
where:
\begin{align}
	F^{(i)}&:=u^{i} - u^{i-1}, \\
    G^{(j)}&:=u^{j-1} - \left(\lambda_{j}\right)u^{j}.
\end{align}
To analyze the structure of the matrix \(K\), we introduce the following integrals:
\begin{align}
I_{ij}(z):=\int_{1}^{z^{2}}F^{(i)}G^{(j)}\,du,\qquad \widehat{\mu}_{k}^{(j)}=I_{ij}(x)+s^{2}I_{ij}(y)-\frac{\Delta}{p^{2}}I_{ij}(0).
\end{align}
By making a change of variables \(u=w^{2}\), the integrals \(I_{ij}(z)\) can be represented by the following expression:
\begin{align}
	I_{ij}(z)= -2(2i-1) \int_{1}^{z} \Gamma_{i}(w) \Gamma_{j}^{\prime}(w) \, dw,\quad \Gamma_{i}(w):= \frac{1}{2i-1}\left[w^{2i-1}-w^{2i+1}\right] , \label{mm:I_ij_def_Gamma}
\end{align}
so that:
\begin{align}
K_{ij}=-2(2i-1)\Bigg[\int_{1}^{x}\Gamma_i(w)\Gamma_j'(w)\,dw+s^{2}\int_{1}^{y}\Gamma_i(w)\Gamma_j'(w)\,dw+\frac{\Delta}{p^{2}}\int_{0}^{1}\Gamma_i(w)\Gamma_j'(w)\,dw\Bigg].
\end{align}
The factor \(-2(2i-1)\) in the above formula can be removed by defining:
\begin{align}
	M_{ij}=-\frac{1}{2(2i-1)}K_{ij},\label{mm:M_K_relation}
\end{align}
so that:
\begin{align}
	\det K = C(n) \det M,
\end{align}
where:
\begin{align}
	C(n):=\prod_{j=1}^{n-1}\big[-2(2j-1)\big] =(-1)^{n-1}\frac{\Gamma(2n-1)}{\Gamma(n)}. \label{mm:C_n_def}
\end{align}
Thus, it is enough to analyze the structure of the matrix \(M\):
\begin{align}
M_{ij}=\int_{1}^{x}\Gamma_i(w)\Gamma_j'(w)\,dw+s^{2}\int_{1}^{y}\Gamma_i(w)\Gamma_j'(w)\,dw+\frac{\Delta}{p^{2}}\int_{0}^{1}\Gamma_i(w)\Gamma_j'(w)\,dw.
\label{mm:M_matrix}
\end{align}
Integration by parts gives:
\begin{align}
	M_{ij} = \omega_{ij} + S_{ij},\label{mm:M_omega_S_split}
\end{align}
where:
\begin{align}
\omega_{ij}:={}&\frac{1}{2}\int_{1}^{x}\left[\Gamma_i(w)\Gamma_j'(w)-\Gamma_i'(w)\Gamma_j(w)\right]\,dw+\frac{s^{2}}{2}\int_{1}^{y}\left[\Gamma_i(w)\Gamma_j'(w)-\Gamma_i'(w)\Gamma_j(w)\right]\,dw \nonumber\\
&+\frac{\Delta}{2p^{2}}\int_{0}^{1}\left[\Gamma_i(w)\Gamma_j'(w)-\Gamma_i'(w)\Gamma_j(w)\right]\,dw,
\label{mm:M_omega_part}\\
S_{ij}:={}&\frac{1}{2}\Gamma_i(x)\Gamma_j(x)+\frac{s^{2}}{2}\Gamma_i(y)\Gamma_j(y).
\label{mm:M_S_part}
\end{align}
We note in the above formulas, that besides separating the matrix \(M\) into symmetric and anti-symmetric parts, we also have a clear separation of \(x\) and \(y\) dependences for each of these parts. Moreover, on the physical subspace, \(\Delta=0\), the parameter \(s \neq 0\) occurs in these formulas only in \(y\)-dependent terms, while for \(s=0\) one has only \(x\)-dependence. This split will be used in the next section to evaluate \(\det M\).

\section{The Pfaffian structure of the polynomial \texorpdfstring{\(f_{n}\)}{fn}}
\label{sec:pf}

In what follows, we denote \(m=n-1\).
The results \eqref{mm:M_omega_part} and \eqref{mm:M_S_part} have the form:
\ba\label{mm:M_general}
M = \Omega+XX^T+YY^T,
\ea
where \(
\Omega\in \operatorname{Mat}_{m}(\mathbb R),
\
\Omega^T=-\Omega
\), \(\Omega_{ij}\equiv \omega_{ij}\)
and
\(
X,Y\in\mathbb R^{m}
\) with \(X_i = \frac{1}{\sqrt{2}}\Gamma_i (x)\) and \(Y_i = \frac{s}{\sqrt{2}}\Gamma_i (y)\).

Define the antisymmetric matrix
\[
C:=XY^T-YX^T,
\qquad
C^T=-C,
\]
and
\ba\label{M+-}
R_\pm:=\Omega\pm \mi \,C.
\ea
Because \(\Omega\) and \(C\) are real,
\[
R:=R_+,
\qquad
R^*=R_-,
\]
where \(^{*}\) denotes entry-wise complex conjugation. Since
\(R^T=-R\), one trivially has
\(
R^\dagger=-R^*.
\)

Introduce also the complex vector
\[
Z:=X+\mi Y,
\quad
Z^*:=X-\mi Y\ \Rightarrow\
Z Z^\dagger
=
XX^T+YY^T-\mi \,C.
\]
Therefore \eqref{mm:M_general} takes the form
\ba\label{R+}
M=R_+ +Z Z^\dagger.
\ea
Because
\[
Z^* Z^T
=
XX^T+YY^T+\mi \,C,
\]
we have equivalently
\ba\label{R-}
R_- = R_+ + Z Z^\dagger - Z^* Z^T.
\ea

Thus the problem reduces to the calculation of the determinant of the matrix \eqref{R+}, \( \det (M) = \det (R_+ + Z Z^\dagger)\). The calculation is a little different for the even and odd cases and we will do it separately for a slightly more general case. We derive and collect all the Pfaffian identities needed below in \ref{app:appa}.

\subsection{The even-dimensional rank-one identity}

Let \(A\in\operatorname{Mat}_{2n}(\mathbb C)\) be antisymmetric,
\(
A^T=-A
\)
and \(a,b\in\mathbb C^{2n}\). Then from the matrix determinant lemma \eqref{matrix-det}
\ba\label{Even1}
\det(A + a b^T) = \det (A) (1 + b^T A^{-1}a).
\ea
Defining \( B := A + a b^T - b a^T \) and applying the rank-two Pfaffian identity \eqref{Pfaffian-rank-2}, we get
\ba\label{Even2}
\pf (B) = \pf (A)(1 + b^T A^{-1}a).
\ea
Because \(\det A  = (\pf A)^2\), we get, by combining \eqref{Even1} and \eqref{Even2}, the rank-one identity
\ba\label{Even3}
\det(A + a b^T) = \pf (A)\pf (B).
\ea

\subsection{The odd-dimensional rank-one identity}

Let \(A\in\operatorname{Mat}_{2n+1}(\mathbb C)\) be antisymmetric,
\(
A^T=-A .
\)
Since \(A\) has odd dimension,
\(\det A=0\), but the matrix determinant lemma \eqref{matrix-det} still can be applied, see footnote \ref{footnote1},
\ba\label{matrix-det-degen}
\det(A+ab^T)
=
\det A+b^T\operatorname{adj}(A)a \equiv b^T\operatorname{adj}(A)a,
\ea
where \(a,b\in\mathbb C^{2n+1}\). Here \(\operatorname{adj}(A)\) is the adjugate matrix, equal to the transposed cofactor matrix of \(A\), i.e. when \(A\) is invertible then \(A^{-1} = \frac{1}{\det(A)}\operatorname{adj}(A)\). Note that \(\operatorname{adj}(A)\) is defined regardless of whether \(A\) is invertible or not. To calculate \(\operatorname{adj}(A)\), for arbitrary \(w\in \mathbb{C}^{2n+1}\), we define the bordered Pfaffian by
\begin{equation}
\pf_w (A)
:=
\pf\begin{pmatrix}
0&w^T\\
-w&A
\end{pmatrix}.\label{bordered_pf_A}
\end{equation}

The bordered Pfaffian introduced above can be seen directly from
the Grassmann representation of \(f_n\). For an odd number of Grassmann
variables, the nonvanishing contribution contains a term linear in the
Grassmann variables. Introducing one additional Grassmann variable
converts this linear term into a quadratic Gaussian form, thus enlarging
the odd-dimensional antisymmetric matrix by one row and one column. The
resulting even-dimensional matrix is precisely the bordered
antisymmetric matrix \eqref{bordered_pf_A}.

Expanding \eqref{bordered_pf_A} along its first row gives, cf. \eqref{Pfaffian:expansion}
\ba\label{Pf_cofactor_id}
\pf_w (A)
=
\sum_{j=1}^{2n+1}
(-1)^{j+1}w_j\,\pf(A_{\widehat{j}}),
\ea
where \(A_{\widehat{j}}\) denotes the antisymmetric matrix obtained from
\(A\) by deleting its \(j\)-th row and column.
It is therefore convenient to introduce the Pfaffian cofactor vector
\[
\pi_j(A):=(-1)^{j+1}\pf(A_{\widehat{j}}).
\]
With this notation,
\[
\pf_w (A)=\pi(A)^T w.
\]
Because the Pfaffian expansion \eqref{Pf_cofactor_id} is valid for any \(w\), then using \(w_j = A_{ij}\) for a fixed \(i\), \eqref{Pf_cofactor_id} becomes the representation of the Pfaffian of a matrix with a duplicated row and column, which is zero, we get
\[
A\,\pi(A)=0.
\]
For the moment, consider a generic antisymmetric matrix \(A\) of rank \(2n\), the kernel of \(A\)
is one-dimensional, so it is spanned by \(\pi(A)\). Since
\[
A\,\operatorname{adj}(A)=
\operatorname{adj}(A)\,A =\det(A)I\equiv 0,
\]
every column and every row of \(\operatorname{adj}(A)\) must be proportional
to \(\pi(A)\) and thus
\[
\operatorname{adj}(A)=c\,\pi(A)\pi(A)^T
\]
for some scalar \(c\). To determine \(c\), consider the diagonal entries. We have
\[
\bigl[\operatorname{adj}(A)\bigr]_{jj}
=
\det(A_{\widehat{j}})
=
\pf(A_{\widehat{j}})^2
=
\pi_j(A)^2
\]
and therefore \(c=1\). Thus
\ba\label{adjA}
\operatorname{adj}(A)=\pi(A)\pi(A)^T.
\ea
As usual, since both sides are polynomial in the entries of \(A\), this identity
continues to hold when \(A\) has rank smaller than \(2n\).

We now use the result \eqref{adjA} in \eqref{matrix-det-degen} to arrive at the analog of \eqref{Even3}:
\ba\label{Odd1}
\det(A+ab^T)
=
b^T\pi(A)\pi(A)^T a
=
\bigl(\pi(A)^T b\bigr)
\bigl(\pi(A)^T a\bigr)
=
\pf_b (A)\,\pf_a (A).
\ea
We can bring \eqref{Odd1} to a form more resembling \eqref{Even3}. Define an upper-triangular matrix, \(S_a\):
\[
S_a=
\begin{pmatrix}
1&a^T\\
0&I_{2n+1}
\end{pmatrix},
\qquad
\det S_a=1.
\]
Then
\[
S_a^T\widehat A_b S_a
=
\begin{pmatrix}
0&b^T\\
-b&A+ab^T-ba^T
\end{pmatrix}
=
\widehat B_b,
\]
where
\[
\widehat A_b=
\begin{pmatrix}
0&b^T\\
-b&A
\end{pmatrix},
\qquad
\widehat B_b=
\begin{pmatrix}
0&b^T\\
-b&B
\end{pmatrix}.
\]
From the Pfaffian congruence identity, we have
\ba\label{shift_sym}
\pf_b (B) = \pf(\widehat B_b)
=
\det(S_a)\pf(\widehat A_b)
\equiv
\pf(\widehat A_b) = \pf_b (A),
\ea
where \( B := A + a b^T - b a^T \) as above. Thus we can identically rewrite \eqref{Odd1} as
\ba\label{Odd2}
\det(A+ab^T)
=
\pf_a (A)\,\pf_b (B).
\ea

\subsection{The polynomial \(f_n\)}

Comparing \eqref{Even1} with \eqref{mm:M_general} and \eqref{R+}, we can identify
\[
A = R_+ \equiv \Omega + iC =: \bar{\omega}^{(s)}  ,\ \ a = Z \equiv \frac{1}{\sqrt{2}}(\Gamma(x) + \mi s \Gamma(y)),\ \ b = Z^* ,\ \ B = R_- \equiv {\omega}^{(s)} ,
\]
where we defined for the future convenience
\ba
\omega^{(s)}_{ij}:= \omega_{ij}-\mi s \widehat{\Gamma}_{ij},\ \ \widehat{\Gamma}_{ij}:= \frac{1}{2} \left[\Gamma_{i}(x)\Gamma_{j}(y) -\Gamma_{i}(y)\Gamma_{j}(x)\right]. \label{pfeven:omega_s}
\ea
Then in the even case, we can immediately apply \eqref{Even3} to get
\begin{align}
\det M =\pf \big(\omega^{(s)} \big) \overline{\pf \big(\omega^{(s)} \big)}. \label{pfeven:f_factorization}
\end{align}

\bigskip

In the odd case, with our identifications, \eqref{Odd1} and \eqref{Odd2} lead to
\ba\label{detR_odd}
\det M=
\pf_{Z^*} (R_+)\,\pf_{Z} (R_+) \equiv \pf_{Z^*} (R_-)\,\pf_{Z} (R_+) = |\pf_{Z}(R)|^2.
\ea
Again, introducing the matrix \(\widehat{\omega}^{(s)}\)
\begin{equation}
\widehat{\omega}^{(s)} =
\begin{pmatrix}
    \omega_{11}  & \cdots & \omega_{1m} & \Gamma_{1}^{(+)} \\
    \omega_{21}  & \cdots & \omega_{2m} & \Gamma_{2}^{(+)} \\
    \vdots & \vdots & \ddots & \vdots  \\
    \omega_{m1}  & \cdots & \omega_{mm} & \Gamma_{m}^{(+)} \\[4pt]
    -\Gamma_{1}^{(+)}  & \cdots & -\Gamma_{m}^{(+)} & 0
\end{pmatrix},\label{pfodd:omega_s}
\end{equation}
where we have defined
\begin{align}
	\Gamma_{j}^{(+)}(x,y):= \Gamma_{j}(x) +\mi s \, \Gamma_{j}(y),
\end{align}
\eqref{detR_odd} can be rewritten as
\begin{align}
2 \det M = \pf \big( \widehat{\omega}^{(s)}\big)  \overline{\pf \big( \widehat{\omega}^{(s)}\big)} .\label{pfodd:f_factorization}
\end{align}

\bigskip

The results of this section can be summarized as
\ba\label{detR}
f_n
=
\frac{(-1)^n p^{2n-2} C(n)}{2n-1} \begin{cases}
\pf \big(\omega^{(s)} \big) \overline{\pf \big(\omega^{(s)} \big)},\ \text{even\ case},\\[4pt]
\frac{1}{2}\pf \big( \widehat{\omega}^{(s)}\big)  \overline{\pf \big( \widehat{\omega}^{(s)}\big)},\ \text{odd\ case}.
\end{cases}
\ea

This completes the proof of the factorization of \(f_n\).

\section{The Pfaffian structure of the polynomial \texorpdfstring{\(g_{n}\)}{gn}}
\label{sec:pfg}

We now consider the \(g_{n}\) polynomial. Setting \(h=0\) in \eqref{mm:hsol_B_matrix} and using \eqref{mm:g_polynomial_def}, the explicit form is:
\begin{equation}
\label{pfg:g_matrix}
g_{n}=\det
\begin{pmatrix}
\zeta_0 & \zeta_1 & \cdots & \zeta_{n-2} & \zeta_{n-1} & \mathsf{b}^{(\mathsf{g})}_{0}\\
\zeta_1 & \zeta_2 & \cdots & \zeta_{n-1} & \zeta_{n}   & \mathsf{b}^{(\mathsf{g})}_{1}\\
\vdots  & \vdots  & \ddots& \vdots      & \vdots      & \vdots\\
\zeta_{n-1} & \zeta_{n} & \cdots & \zeta_{2n-3} & \zeta_{2n-2} & \mathsf{b}^{(\mathsf{g})}_{n-1}\\[1ex]
1 & 1 & \cdots & 1 & 1 & 0
\end{pmatrix},
\end{equation}
where the moments are exactly the same as before (for \(h=0\)):
\begin{align}
	\zeta_{m}(x,y,p,q;0)=\frac{1}{m+1}\left[p^{2} x^{2m+2}+q^{2}y^{2m+2} - 1 \right],
\end{align}
while the border elements \(\mathsf{b}^{(\mathsf{g})}_{\ell}\) have the form:
\begin{align}
	\mathsf{b}^{(\mathsf{g})}_{\ell} = \frac{1}{2 \ell +1}\left[p x^{2 \ell +1} -\mi q y^{2 \ell+1}\right]. \label{pfg:gn_border_h_0}
\end{align}
Due to the polynomial nature of the border elements \(\mathsf{b}^{(\mathsf{g})}_{\ell}\), it is possible, in contrast to the case of \(f_n\), to reduce the dimension of the matrix in \eqref{pfg:g_matrix} only once. Nevertheless, the factorization formulas \eqref{intro:main_factorization} essentially suggest the next step: since both \(g_n\) and \(f_n\) are bordered determinants with the same Hankel block, it is natural to expect this common block to be responsible for the common divisor \(d_n\). We therefore apply to \(g_n\) the same column transformations used for \(f_n\), which produce the same main block after these transformations.

The analysis in this section is parallel to that of the polynomial \(f_{n}\) given in the previous sections. The main difference is in the border structure, which has a more complex polynomial form for the polynomial \(g_{n}\). After the two column operations \eqref{mm:column_op_1} and \eqref{mm:column_op_2} applied to \eqref{pfg:g_matrix}, one obtains:
\begin{equation}
\label{pfg:g_matrix_after_column_ops}
g_{n}=(-1)^{n}p^{2n-1}
\det \begin{pmatrix}
\widehat{\mu}^{(1)}_{0}  & \widehat{\mu}_{1}^{(2)} & \cdots & \widehat{\mu}^{(n-1)}_{n-2} & \alpha_{1}\\
\widehat{\mu}^{(1)}_{1}  & \widehat{\mu}_{2}^{(2)} & \cdots  & \widehat{\mu}^{(n-1)}_{n-1}   & \alpha_{2}\\
\vdots  & \ddots & \vdots      & \vdots      \\
\widehat{\mu}^{(1)}_{n-2} & \widehat{\mu}_{n-1}^{(2)} & \cdots& \widehat{\mu}^{(n-1)}_{2n-4} & \alpha_{n-1}\\[1ex]
\beta_{1} & \beta_{2} & \cdots & \beta_{n-1} & \gamma
\end{pmatrix}.
\end{equation}
Here, the elements \(\widehat{\mu}^{(j)}_{i+j-2}\) of the main block  are exactly those of the matrix \(K\) for the polynomial \(f_{n}\) \eqref{mm:f_n_K_matrix}, and the explicit expressions for the borders \(\alpha_k, \beta_k\) and \(\gamma\) are:
\begin{align}
&\alpha_{k}:= \int_{1}^{x^{2}}u^{k+n-2}(u-1)\,du + s^{2}\int_{1}^{y^{2}}u^{k+n-2}(u-1)\,du + \frac{\Delta}{p^{2}}\int_{0}^{1}u^{k+n-2}(u-1)\,du, \label{pfg:alpha_k}\\
&\beta_{k}:=\frac{1}{2k-1}\left[x^{2k-1}(1-x^{2})-\mi s y^{2k-1}(1-y^{2})\right], \label{pfg:beta_k}\\
&\gamma:=\frac{1}{2n-1}\left[x^{2n-1}-\mi s y^{2n-1}\right]. \label{pfg:gamma}
\end{align}
The determinant structure of \eqref{pfg:g_matrix_after_column_ops} shows that the column transformations reduce the determinant for \(g_{n}\) to a bordered determinant with the main block corresponding to \(f_{n}\) (cf. \eqref{mm:f_n_K_matrix}). Factoring out numerical constants \(\left(-2 \left[2j-1 \right]\right)\) from the rows \(j=1,\dots,n-1\) in \eqref{pfg:g_matrix_after_column_ops}, we can represent \(g_{n}\) in terms of the same matrix \(M\) in \eqref{mm:M_matrix} as follows:
\begin{align}
\label{pfg:gn_M_matrix}
g_n
=  (-1)^n p^{2n-1} C(n) \det M^{(g)},
\end{align}
where:
\begin{align}
M^{(g)} =
\begin{pmatrix}
M_{11}  & M_{12} & \cdots & M_{1,n-1} & \rho_{1}\\
M_{21}  & M_{22} & \cdots  & M_{2,n-1}   & \rho_{2}\\
\vdots  & \ddots & \vdots      & \vdots      \\
M_{n-1,1} & M_{n-1,2} & \cdots& M_{n-1,n-1} & \rho_{n-1}\\[1ex]
\beta_{1} & \beta_{2} & \cdots & \beta_{n-1} & \gamma
\end{pmatrix}.\label{pfg:gn_M_matrix_explicit}
\end{align}
Here we have denoted:
\begin{align}
\rho_{k}:= -\frac{1}{2\left(2k-1 \right)} \alpha_{k}, \label{pfg:rho_def}
\end{align}
and \(M_{ij}\) was defined in \eqref{mm:M_K_relation}.
Using the explicit expressions \eqref{pfg:alpha_k}-\eqref{pfg:gamma} for \(\alpha_{k}, \beta_{k}, \gamma\) and for \(\rho_{k}\) in \eqref{pfg:rho_def}, we obtain the expressions for  \(\rho_{k}, \beta_{k}\) in terms of \(\Gamma_{i}(w)\) defined in \eqref{mm:I_ij_def_Gamma}:
\begin{align}
&\rho_{k} = \int_{1}^{x}\Gamma_{k}(w)w^{2n-2}\,dw +s^{2}\int_{1}^{y}\Gamma_{k}(w)w^{2n-2}\,dw + \frac{\Delta}{p^{2}}\int_{0}^{1}\Gamma_{k}(w)w^{2n-2}\,dw, \label{pfg:rho_explicit}\\
&\beta_{k} = \Gamma_{k}(x)-\mi s\Gamma_{k}(y). \label{pfg:beta_explicit}
\end{align}

To find \(\det M^{(g)}\), we adopt the method of Section \ref{sec:pf}. We use exactly the same notation as in Section \ref{sec:pf}, i.e.,
\ba\label{Identifications}
A = R_+ \equiv \Omega + iC =: \bar{\omega}^{(s)}  ,\ \ a = Z \equiv \frac{1}{\sqrt{2}}(\Gamma(x) + \mi s \Gamma(y)),\ \ b = Z^* ,\ \ B = R_- \equiv {\omega}^{(s)}
\ea
and, in addition, define \(\lambda = \sqrt{2}\). In these notations the matrix \eqref{pfg:gn_M_matrix_explicit} takes the form
\[
M^{(g)}= \begin{pmatrix}
M&\rho\\
\beta^T&\gamma
\end{pmatrix} =
\begin{pmatrix}
A+ab^T&\rho\\
\lambda b^T&\gamma
\end{pmatrix}.
\]
As before, we have to separately analyze the even and odd cases. The special form of the last row will prove to be crucial for simplifying the calculations.

\paragraph{The even case: \(m=2k\)}
For invertible \(M\) and \(\gamma\ne 0\), applying the matrix determinant lemma \eqref{matrix-det} twice (or using the Schur complement directly) gives
\ba\label{M_g_even_1}
\det M^{(g)} = \det\begin{pmatrix}M&\rho\\ \lambda b^T&\gamma\end{pmatrix}
=\det M\,(\gamma-\lambda b^T M^{-1}\rho)
=\gamma\det M-\lambda b^T\operatorname{adj}(M)\rho.
\ea
The last equality holds for singular \(M\) and \(\gamma\) as well, by polynomial continuation. Now introduce the one-parameter family
\ba\label{Mt}
M(t)=A+(a+t\rho)b^T=M+t\rho b^T.
\ea
Then
\[
\det M(0)=\det M,\qquad
\left.\frac{d}{dt}\det M(t)\right|_{t=0}
=b^T\operatorname{adj}(M)\rho,
\]
so from \eqref{M_g_even_1}
\ba\label{detMt}
\det M^{(g)}
=\gamma\det M-\lambda\left.\frac{d}{dt}\det M(t)\right|_{t=0}.
\ea

For every \(t\), the even rank-one identity \eqref{Even3} gives
\[
\det M(t)=\pf(A)\pf(B(t)),
\qquad
B(t)=B+t(\rho b^T-b\rho^T),
\]
leading to
\[
\det M^{(g)}
=\pf(A)\left[
\gamma\pf(B)
-\lambda\left.\frac{d}{dt}\pf(B(t))\right|_{t=0}
\right].
\]
For invertible \(B\), the Pfaffian block formula and the rank-two Pfaffian identity \eqref{Pfaffian-rank-2} give\footnote{As usual, we first consider the invertible case and the resulting polynomial identity then extends to singular matrices by continuity.}
\[
\begin{aligned}
\pf\begin{pmatrix}
B&\rho&\lambda b\\
-\rho^T&0&\gamma\\
-\lambda b^T&-\gamma&0
\end{pmatrix}
=\pf(B)\bigl(\gamma+\lambda\rho^TB^{-1}b\bigr)
=\gamma\pf(B)-\lambda\left.\frac{d}{dt}\pf B(t)\right|_{t=0}.
\end{aligned}
\]
Thus we finally get
\[
\det M^{(g)}
=\pf(A)\,
\pf\begin{pmatrix}
B&\rho&\lambda b\\
-\rho^T&0&\gamma\\
-\lambda b^T&-\gamma&0
\end{pmatrix}.
\]
Using the definitions \eqref{Identifications}, one immediately obtains
\begin{align}
	\det M^{(g)} =  \overline{\pf \big(\omega^{(s)} \big)} \pf \big(\Omega^{(e)}\big),
\end{align}
where \(\omega^{(s)}\) is given in \eqref{pfeven:omega_s}, and the \((m+2) \times (m+2)\) antisymmetric matrix \(\Omega^{(e)}\) is defined by:
\begin{align}
    \Omega^{(e)}
    =
    \left(
    \begin{array}{cccc|cc}
    0 & \omega^{(s)}_{12} & \cdots & \omega^{(s)}_{1m} & \rho_1 & \beta_1 \\
    -\omega^{(s)}_{12} & 0 & \cdots & \omega^{(s)}_{2m} & \rho_2 & \beta_2 \\
    \vdots & \vdots & \ddots & \vdots & \vdots & \vdots \\
    -\omega^{(s)}_{1m} & -\omega^{(s)}_{2m} & \cdots & 0 & \rho_m & \beta_m \\[4pt]
    \hline
    -\rho_1 & -\rho_2 & \cdots & -\rho_m & 0 & \gamma \\
    -\beta_1 & -\beta_2 & \cdots & -\beta_m & -\gamma & 0
    \end{array}
    \right).
\end{align}
The Yamazaki-Hori solution to the Ernst equation for this case has the form:
\begin{align}\label{xi_n_1}
	\xi_{n}  = p \left( 2n-1 \right)\frac{ \pf \big( \Omega^{(e)} \big)}{\pf \big( {\omega}^{(s)}\big)}.
\end{align}

Alternatively, using the rank-two Pfaffian identity \eqref{Pfaffian-rank-2}, one can show that:
\begin{align}
	\det M^{(g)} &=  \overline{\pf \big(\omega^{(s)} \big)} \pf \big(\omega^{(s)} \big) \left[ \gamma  + {\rho}^{T}\big(\omega^{(s)} \big)^{-1} {\beta} \right],\label{pfg:M_g_second_fact_id}
\end{align}
from which we find:
\begin{align}
	g_{n} = p \left( 2n-1 \right) f_{n} \left[ \gamma  + {\rho}^{T}\big(\omega^{(s)} \big)^{-1} {\beta} \right].
\end{align}
This shows that the Yamazaki-Hori solution of the Ernst equation can also be recast in terms of the inverse matrix \(\big(\omega^{(s)} \big)^{-1}\) :
\begin{align}
	\xi_{n} = \frac{g_{n}}{f_{n}}=p \left( 2n-1 \right)\left[ \gamma  + {\rho}^{T}\big(\omega^{(s)} \big)^{-1} {\beta} \right] . \label{pfg:Ernst_factor}
\end{align}

\paragraph{The odd case: \(m=2k+1\)}
The result \eqref{detMt} does not depend on \(m\) being even or odd. But to calculate the determinant of \(M(t)\) in the odd case, we have to use \eqref{Odd2}:
\[
\det M(t)=\pf_{a+t\rho}(A)\,\pf_b(B(t)),
\qquad
B(t)=A+(a+t\rho)b^T-b(a+t\rho)^T.
\]
From the bordered-Pfaffian shift identity \eqref{shift_sym} we have \(\pf_b(B(t))=\pf_b(A)\), while \(\pf_w(A)\) is linear in \(w\), see \eqref{Pf_cofactor_id}. Hence
\[
\det M(t)=\bigl(\pf_a(A)+t\pf_\rho(A)\bigr)\pf_b(A),
\]
and therefore from \eqref{detMt}
\[
\begin{aligned}
\det M^{(g)} =\gamma\det M-\lambda\left.\frac{d}{dt}\det M(t)\right|_{t=0}
=\pf_b(A)\bigl(\gamma\pf_a(A)-\lambda\pf_\rho(A)\bigr)
=\lambda\pf_b(A)\pf_{\frac{\gamma}{\lambda}a-\rho}(A).
\end{aligned}
\]
With \(\lambda=\sqrt2\) and \(\overline{\beta}=\lambda \overline{b}\equiv\lambda a\), the new border is
\[
\frac{\gamma}{\lambda}a-\rho
=\frac{\gamma}{2}\,\overline{\beta}-\rho
=:\sigma.
\]

The remaining prefactor is related to \(\overline{\widehat{\omega}^{(s)}}\), defined in \eqref{pfodd:omega_s}, by the same unit-determinant congruence used in the odd-\(f_n\) calculation
\[
S=\begin{pmatrix}I&-\dfrac{a}{2\lambda}\\0&1\end{pmatrix},\qquad
S\begin{pmatrix}A&\lambda b\\-\lambda b^T&0\end{pmatrix}S^T
=\begin{pmatrix}\omega&\lambda b\\-\lambda b^T&0\end{pmatrix}
=\overline{\widehat{\omega}^{(s)}},\qquad \det S=1
\]
and we have
\[
\overline{\pf\!\left(\widehat{\omega}^{(s)}\right)}
=\lambda\pf_b(A).
\]
Combining these results and using the same odd-dimensional border-ordering convention as in Section \ref{sec:pf}, we finally get
\begin{align}
		\det M^{(g)} =   \overline{\pf \big( \widehat{\omega}^{(s)}\big)} \pf \big( \Omega^{(o)} \big) \label{pfgodd:M_g_odd_factorization}
\end{align}
where  the antisymmetric matrix \(\Omega^{(o)}\) has the form:
\begin{align}
\Omega^{(o)}
=
\left(
\begin{array}{cccc|c}
0 & \omega^{(-s)}_{12} & \cdots & \omega^{(-s)}_{1m} & \sigma_1 \\
-\omega^{(-s)}_{12} & 0 & \cdots & \omega^{(-s)}_{2m} & \sigma_2 \\
\vdots & \vdots & \ddots & \vdots & \vdots \\
-\omega^{(-s)}_{1m} & -\omega^{(-s)}_{2m} & \cdots & 0 & \sigma_m \\[4pt]
\hline
-\sigma_1 & -\sigma_2 & \cdots & -\sigma_m & 0
\end{array}
\right), \quad \sigma_{j} =\frac{\gamma}{2}\overline{\beta_{j}} - \rho_{j}.
\end{align}
The Yamazaki-Hori solution to the Ernst equation for this case has the form:
\begin{align}\label{xi_n_2}
	\xi_{n}  = 2p \left( 2n-1 \right) \frac{ \pf \big( \Omega^{(o)} \big)}{\pf \big( \widehat{\omega}^{(s)}\big)}.
\end{align}

Thus, we have shown that for both even and odd cases, the Yamazaki-Hori solution is given by the ratio of two Pfaffians. Collecting the above results, for \(m=n-1=2k\) we have:
\begin{align}
	f_{n} &= C^{(f)} \, \overline{\pf \big(\omega^{(s)}\big)} \pf \big(\omega^{(s)}\big), \quad C^{(f)}= \frac{(-1)^n p^{2n-2} C(n)}{2n-1}, \\
	g_{n} &= C^{(g)}\, \overline{\pf \big(\omega^{(s)}\big)} \pf \big(\Omega^{(e)}\big), \quad C^{(g)}= (-1)^n p^{2n-1} C(n).
\end{align}
For \(m=n-1=2k+1\), the corresponding formulas are:
\begin{align}
	f_{n} &= \widetilde{C}^{(f)}\,  \overline{\pf \big(\widehat{\omega}^{(s)}\big)} \pf \big(\widehat{\omega}^{(s)}\big), \quad \widetilde{C}^{(f)} = \frac{(-1)^n p^{2n-2} C(n)}{2(2n-1)},\\
	g_{n}&= \widetilde{C}^{(g)} \, \overline{\pf \big(\widehat{\omega}^{(s)}\big)} \pf \big(\Omega^{(o)}\big), \quad \widetilde{C}^{(g)}=(-1)^n p^{2n-1}C(n).
\end{align}
In the formulas, \(C(n)\) is defined in \eqref{mm:C_n_def}.

\section{Polynomial factors and the greatest common divisor}
\label{pfg:polynomial_gcd}

We now show that the Pfaffian factors obtained above are polynomials in the independent variables \(p,q,x,y\), and identify the common factor with the greatest common divisor established in \cite{Melikyan:2026ux}. Let \(\mathcal{R}:=\mathbb{C}[p,q,x,y]\), and define:
\begin{align}
P_n &:=
\begin{cases}
p^{n-1}\pf\big(\omega^{(s)}\big), & m=n-1\text{ even},\\[4pt]
p^{n-1}\pf\big(\widehat{\omega}^{(s)}\big), & m=n-1\text{ odd},
\end{cases}
\label{pfg:P_def}\\
Q_n &:=
\begin{cases}
p^n\pf\big(\Omega^{(e)}\big), & m=n-1\text{ even},\\[4pt]
p^n\pf\big(\Omega^{(o)}\big), & m=n-1\text{ odd}.
\end{cases}
\label{pfg:Q_def}
\end{align}
Recall that \(s=q/p\) and \(\Delta=p^2+q^2-1\). The integral representations \eqref{mm:M_omega_part} and \eqref{pfg:rho_explicit}, together with the definitions of the border elements, imply:
\begin{align}
p^2\omega_{ij},\quad p^2\omega_{ij}^{(\pm s)},\quad p^2\rho_i,\quad p\beta_i,\quad p\gamma,\quad p^2\sigma_i,\quad p\Gamma_i^{(+)}\in\mathcal{R}.
\label{pfg:polynomial_entries}
\end{align}
\begin{lemma}
	\(P_n\) and \(Q_n\) belong to \(\mathcal R\)
\end{lemma}
\begin{proof}
To prove this, we only use the Pfaffian identity:
\begin{align}
\pf(DAD^T)=\det(D)\pf(A),
\label{pfg:pf_congruence}
\end{align}
valid for any antisymmetric matrix \(A\).
We consider the even and odd cases separately.

\paragraph{Even case: \(m=n-1=2k\)}

First, consider \(P_n=p^m\pf(\omega^{(s)})\). Define the antisymmetric matrix
\begin{align}
\widetilde{\omega}^{(s)}:=p^2\omega^{(s)}.
\end{align}
All entries of \(\widetilde{\omega}^{(s)}\) belong to \(\mathcal R\). Moreover, since \(m=2k\), applying \eqref{pfg:pf_congruence} with \(D=pI_m\) gives
\begin{align}
\pf(\widetilde{\omega}^{(s)})
=\det(pI_m)\pf(\omega^{(s)})
=p^m\pf(\omega^{(s)})=P_n.
\end{align}
Thus, \(P_n\in\mathcal R\).

Next, consider the matrix \(\Omega^{(e)}\). Multiplying all its entries by \(p^2\), we obtain
\begin{align}
p^2\Omega^{(e)}
=
\begin{pmatrix}
p^2\omega^{(s)} & p^2\rho & p^2\beta\\
-p^2\rho^T & 0 & p^2\gamma\\
-p^2\beta^T & -p^2\gamma & 0
\end{pmatrix}.
\end{align}
All entries of this matrix belong to \(\mathcal R\). Furthermore, since \(p\beta_i,p\gamma\in\mathcal R\), we can divide its last row and column by \(p\), obtaining another antisymmetric matrix with polynomial entries:
\begin{align}
\widetilde{\Omega}^{(e)}
:=
\begin{pmatrix}
p^2\omega^{(s)} & p^2\rho & p\beta\\
-p^2\rho^T & 0 & p\gamma\\
-p\beta^T & -p\gamma & 0
\end{pmatrix}.
\label{pfg:Omega_e_polynomial}
\end{align}
Notice that this matrix is related to \(\Omega^{(e)}\) by
\begin{align}
\widetilde{\Omega}^{(e)}
=D_e\Omega^{(e)}D_e^T,
\qquad
D_e:=\operatorname{diag}(pI_m,p,1).
\end{align}
Since \(\det D_e=p^{m+1}=p^n\), the identity \eqref{pfg:pf_congruence} gives
\begin{align}
\pf(\widetilde{\Omega}^{(e)})
=p^n\pf(\Omega^{(e)})=Q_n.
\end{align}
Consequently, \(Q_n\in\mathcal R\).

\paragraph{Odd case: \(m=n-1=2k+1\)}

In this case, \(P_n=p^m\pf(\widehat{\omega}^{(s)})\). Multiplying all entries of the bordered antisymmetric matrix \(\widehat{\omega}^{(s)}\) by \(p^2\), we obtain
\begin{align}
p^2\widehat{\omega}^{(s)}
=
\begin{pmatrix}
p^2\omega & p^2\Gamma^{(+)}\\
-p^2(\Gamma^{(+)})^T & 0
\end{pmatrix}.
\end{align}
Since \(p\Gamma_i^{(+)}\in\mathcal R\), we can again divide the last row and column by \(p\), obtaining the polynomial antisymmetric matrix
\begin{align}
\widetilde{\widehat{\omega}}^{(s)}
:=
\begin{pmatrix}
p^2\omega & p\Gamma^{(+)}\\
-p(\Gamma^{(+)})^T & 0
\end{pmatrix}.
\label{pfg:omega_hat_polynomial}
\end{align}
This matrix satisfies
\begin{align}
\widetilde{\widehat{\omega}}^{(s)}
=D_o\widehat{\omega}^{(s)}D_o^T,
\qquad
D_o:=\operatorname{diag}(pI_m,1).
\end{align}
Using \(\det D_o=p^m=p^{n-1}\), we obtain
\begin{align}
\pf(\widetilde{\widehat{\omega}}^{(s)})
=p^{n-1}\pf(\widehat{\omega}^{(s)})=P_n.
\end{align}
Hence \(P_n\in\mathcal R\).

Finally, consider \(Q_n=p^n\pf(\Omega^{(o)})\). Here it is enough to multiply all entries of \(\Omega^{(o)}\) by \(p^2\). Define
\begin{align}
\widetilde{\Omega}^{(o)}
:=p^2\Omega^{(o)}
=
\begin{pmatrix}
p^2\omega^{(-s)} & p^2\sigma\\
-p^2\sigma^T & 0
\end{pmatrix}.
\label{pfg:Omega_o_polynomial}
\end{align}
All entries of this matrix belong to \(\mathcal R\). Since \(\Omega^{(o)}\) has order \(m+1=n\), which is even, applying \eqref{pfg:pf_congruence} with \(D=pI_n\) gives
\begin{align}
\pf(\widetilde{\Omega}^{(o)})
=\det(pI_n)\pf(\Omega^{(o)})
=p^n\pf(\Omega^{(o)})=Q_n.
\end{align}
Thus, \(Q_n\in\mathcal R\).

Thus, for both even and odd \(m=n-1\), we have expressed \(P_n\) and \(Q_n\) as Pfaffians of antisymmetric matrices with polynomial entries. Hence,
\begin{align}
P_n,Q_n\in\mathcal R.
\label{pfg:PQ_polynomial}
\end{align}
\end{proof}
\noindent The factorizations obtained in the preceding sections can therefore be rewritten as identities between polynomials:
\begin{align}
f_n&=\frac{(-1)^nC(n)}{2n-1}
\begin{cases}
P_n\overline{P_n},&m\text{ even},\\[4pt]
\frac{1}{2}P_n\overline{P_n},&m\text{ odd},
\end{cases}
\label{pfg:PQ_factor_f}\\
g_n&=(-1)^nC(n)\overline{P_n}Q_n.
\label{pfg:PQ_factor_g}
\end{align}
We stress that these identities hold in the full ring \(\mathcal{R}\), without imposing the physical constraint \(p^2+q^2=1\).

We now relate the greatest common divisor \(d_n=\gcd_{\mathcal{R}}(f_n,g_n)\) in \eqref{intro:main_factorization} to \(\overline{P_n}\) in the above factorization formulas.

\begin{lemma}
        \(d_{n}\sim\overline{P_n}\).
\end{lemma}
\begin{proof}
It was shown in \cite{Melikyan:2026ux}, as discussed in the Introduction, that \(f_n=\mathcal{K}(n)d_n\overline{d_n}\), where \(\mathcal{K}(n)\) is a nonzero constant depending only on \(n\), and \(d_n=\gcd_{\mathcal{R}}(f_n,g_n)\) (see \eqref{intro:main_factorization}). It follows from \eqref{pfg:PQ_factor_f} and \eqref{pfg:PQ_factor_g} that \(\overline{P_n}\) divides both \(f_n\) and \(g_n\). Since \(\overline{P_n}\) is a common divisor, there exists \(H_n\in\mathcal R\) such that:
\begin{align}
d_n=\overline{P_n}H_n.
\end{align}
Comparing the two expressions for \(f_n\), we obtain:
\begin{align}
H_n\overline{H_n}\in\mathbb{C}^{\times}.
\end{align}
Since \(H_n\) is a polynomial, this implies that \(H_n\) is a nonzero constant. Thus:
\begin{align}
d_n\sim\overline{P_n}.
\label{pfg:gcd_identification}
\end{align}
\end{proof}
\noindent As a consequence, we have the following:
\begin{corollary}
    The factors \(P_n,Q_n\) are coprime.
\end{corollary}
\begin{proof}
Indeed, since \(\mathcal{R}\) is a unique factorization domain, the factorization formulas \eqref{pfg:PQ_factor_f} and \eqref{pfg:PQ_factor_g} imply:
\begin{align}
\gcd_{\mathcal{R}}(f_n,g_n)\sim\overline{P_n}\,\gcd_{\mathcal{R}}(P_n,Q_n).
\end{align}
Comparing this relation with \eqref{pfg:gcd_identification}, we see that
\begin{align}
\gcd_{\mathcal{R}}(P_n,Q_n)=1.
\label{pfg:PQ_coprime}
\end{align}
\end{proof}
\noindent Thus the polynomial Pfaffian factor \(\overline{P_n}\) is precisely the greatest common divisor of \(f_n\) and \(g_n\), up to a nonzero constant, and the remaining factors \(P_n,Q_n\) are coprime.\footnote{For \(n=2,3\), direct evaluation gives \(\overline{P_2}=-6d_2\) and \(\overline{P_3}=30d_3\), in agreement with the explicit greatest common divisors given in Appendix A of \cite{Melikyan:2026ux}.}

\bigskip
Note that in the proof we work in the full polynomial ring \(\mathcal{R}:=\mathbb{C}[p,q,x,y]\), without imposing the physical constraint \(\Delta=0\). In general, upon restriction to the physical subspace, the greatest common divisor may increase. This, however, does not affect the proof of Nakamura's conjecture in \cite{MP2026nc}, where the Toda \(\tau\)-functions are identified with \(P_n\) and \(Q_n\), up to nonzero constant factors.\footnote{Fukuyama, Kamimura and Yu \cite{Fukuyama:1995cm} expressed the Toda polynomials as minors of an enlarged Wronskian determinant and proved the first two bilinear Ernst equations in \cite{Nakamura:1991sv} for general \(p,q\) satisfying \(p^2+q^2=1\), while establishing the remaining two equations only in the Weyl case \(p=1,q=0\). Fukuyama and Koizumi \cite{Fukuyama:2011ma} subsequently proved the highest and lowest orders in a Laurent expansion in the rotation parameter. Denoting the Toda polynomials of \cite{Fukuyama:1995cm} by \(F_n,G_n\), we find, on the physical subspace \(\Delta=0\), that \((f_2,g_2)=\frac{\overline{P_2}}{6}(F_2,G_2)=-d_2(F_2,G_2)\) and \((f_3,g_3)=-\frac{\overline{P_3}}{120}(F_3,G_3)=-\frac{1}{4}d_3(F_3,G_3)\).}

\section{Conclusion}

We have established an explicit Pfaffian representation of the Yamazaki-Hori solutions to the Ernst equation for arbitrary \(n\). In particular, we have shown that the polynomials \(f_n\) and \(g_n\) admit a factorization in terms of two polynomial Pfaffians \(P_n\) and \(Q_n\), and that \(d_n\sim\overline{P_n}\), where \(d_n\) is their greatest common divisor in \(\mathbb{C}[p,q,x,y]\). This provides a constructive realization of the factorization established in \cite{Melikyan:2026ux} and an analytical proof of the Pfaffian structure conjectured by Nakamura and Ohta in \cite{Nakamura:1991sv}.

The main consequence of the Pfaffian representation obtained here is the proof of Nakamura's conjecture, given in \cite{MP2026nc}. In that work, we establish the following relations between the Toda \(\tau\)-functions and the Pfaffians \(P_n\) and \(Q_n\):
\begin{align}
\tau_{n-1}(\Phi')&\sim P_n
=p^{n-1}
\begin{cases}
\pf(\omega^{(s)}),&n-1\text{ even},\\[4pt]
\pf(\widehat{\omega}^{(s)}),&n-1\text{ odd},
\end{cases}\\
\tau_n(\Phi)&\sim Q_n
=p^n
\begin{cases}
\pf(\Omega^{(e)}),&n-1\text{ even},\\[4pt]
\pf(\Omega^{(o)}),&n-1\text{ odd}.
\end{cases}
\end{align}
The proof is based on a transformation between the polynomial sequences entering the Toda determinants and the polynomials \(\Gamma_j\) introduced in this paper. This transformation allows the elements of the antisymmetric matrices to be expressed as alternating sums of products of the Toda polynomials. An extension of the generalized Cusick identity \cite{Vein:2013kv} then relates the corresponding Pfaffians to the block-Hankel determinants defining the Toda \(\tau\)-functions. Thus, the Pfaffian representation provides the explicit algebraic connection needed to establish Nakamura's conjecture.

As another potential application of the results of this paper, we briefly mention the possible a deeper symmetry or geometric interpretation of the correspondence. The Pfaffian structure of \(\xi_n\), \eqref{xi_n_1} or \eqref{xi_n_2}, suggests that it might have an interpretation in terms of superspace. E.g. in the even case, if \(P\) is a polarization of \(R\) in \eqref{M+-}
\[
P^T R P
=
\begin{pmatrix}
0&Q\\
-Q^T&0
\end{pmatrix},
\qquad
\pf(R)=\det Q ,
\]
then \(\xi_n\) has a representation, schematically given by
\ba\label{QQ}
\begin{aligned}
\xi_n \sim
\int&
D\bar bDb\,D\bar cDc\,D\bar\psi\,D\psi\,D\bar\xi\,D\xi\\
&\times
\exp\left[
-\bar b^TQb
-\bar c^TQ^*c
+\bar\xi^T Q^*\xi
+\bar\psi^T Q \psi + \frac{1}{2}\Psi^T \Delta_Q \Psi
\right].
\end{aligned}
\ea
Here \(b,\bar{b},c,\bar{c}\) are bosonic and \(\xi,\bar\xi,\psi,\bar\psi\) fermionic coordinates of the superspace. One can see that \(\Delta_Q\) is the only term that breaks \(Q \leftrightarrow Q^*\) symmetry and exactly this term is responsible for \(g_n \ne f_n\). We are planning to report on the progress in this direction elsewhere.

\appendix
\section{Grassmann derivation of Pfaffian identities}
\label{app:appa}

Let $M\in \mathrm{Mat}_{2n}$ and
\(
M^T=-M.
\)\footnote{Whenever an inverse appears, we also assume the corresponding matrix to be nonsingular. The resulting identities without inverses are polynomial identities in the matrix entries and therefore extend to singular matrices by continuity.\label{footnote1}}
To fix the conventions, we begin with the standard Grassmann representation of the determinant \cite{Berezin-1987is,zinnjustin-2021qft}.
Thus, introduce independent Grassmann variables
\[
\psi_i,\ \bar\psi_i,\ i=1,\dots,2n
\]
and define the Berezin measure (the ordering is important)
\[
D(\bar\psi,\psi)
=
d\bar\psi_{2n}\,d\psi_{2n}\cdots
d\bar\psi_1\,d\psi_1,
\]
with the usual rules for the Berezin integral
\[
\int d\psi_i\,1=0,\qquad
\int d\psi_i\,\psi_i=1,
\]
or, in general,
\[
\int d\psi_s d\psi_{s-1}\cdots d\psi_1\,\psi_{i_1}\psi_{i_2}\cdots\psi_{i_k}=\delta_{sk}\epsilon_{i_1 i_2 \cdots i_k},
\]
and similarly for $\bar\psi_i$.
Then
\[
\det M
=
\int D(\bar\psi,\psi)\,
\exp\!\left(\bar\psi^TM\psi\right)
\]
and
\ba\label{1-rank}
(M^{-1})_{ji}\det M
=
\int D(\bar\psi,\psi)\,
\exp\!\left(\bar\psi^TM\psi\right)\bar\psi_i \psi_j .
\ea
This immediately leads to the matrix determinant lemma
\ba\label{matrix-det}
\det (M + X Y^T) = \int D(\bar\psi,\psi)\,
\exp\!\left(\bar\psi^T (M + X Y^T)\psi\right) \nonumber\\
=\int D(\bar\psi,\psi)\,
\exp\!\left(\bar\psi^T M\psi\right) (1 + \bar\psi^T  X Y^T\psi )
\stackrel{\eqref{1-rank}}{=} \det M\, (1 + Y^T M^{-1} X),
\ea
where $X,Y \in \mathbb{C}^{2n}$.\footnote{The determinant identities are dimension-independent; for the Pfaffian factorization below, we restrict to even dimension \(2n\).}

Now, introduce two real Grassmann vectors $\xi,\eta$ by
\[
\psi=\frac{\xi+i\eta}{\sqrt2},
\qquad
\bar\psi=\frac{\xi-i\eta}{\sqrt2}.
\]
The quadratic form becomes
\begin{align}\label{real}
\bar\psi^TM\psi
=
\frac12(\xi-i\eta)^TM(\xi+i\eta)
=
\frac12\xi^TM\xi
+
\frac12\eta^TM\eta,
\end{align}
where we used, since $M^T=-M$,
\(
\eta^TM\xi
=
-\xi^TM^T\eta
=
\xi^TM\eta
\).

Also, for each pair,
\[
d\bar\psi_i\,d\psi_i
=
-i\,d\xi_i\,d\eta_i.
\]
The sign could be seen, e.g. from the evaluation of \( \int d\bar\psi d \psi\, \bar\psi\psi = -1 = \int (-id\xi d\eta)\, (i\xi\eta)\).
Hence
\ba\label{measure}
D(\bar\psi,\psi)
=
(-1)^n\,
d\xi_{2n}d\eta_{2n}\cdots
d\xi_1d\eta_1.
\ea
Define
\[
D\xi=d\xi_{2n}\cdots d\xi_1,
\qquad
D\eta=d\eta_{2n}\cdots d\eta_1.
\]
Then, after a rearrangement, \eqref{measure} becomes
\[
D(\bar\psi,\psi)=D\xi\,D\eta.
\]
Consequently, combining with \eqref{real}, we have
\[
\det M
=
\int
D\xi\,D\eta\,
\exp\!\left(
\frac12\xi^TM\xi
+
\frac12\eta^TM\eta
\right),
\]
which trivially factorizes
\[
\det M
=
\left(
\int D\xi\,
e^{\frac12\xi^TM\xi}
\right)
\left(
\int D\eta\,
e^{\frac12\eta^TM\eta}
\right).
\]
Because the result of each integration is polynomial in the entries of $M$, this motivates the definition
\ba\label{Pfaffian}
\pf M
=
\int D\xi\,
\exp\!\left(
\frac12\xi^TM\xi
\right),
\ea
for any \(2n\times2n\) antisymmetric matrix $M$, resulting in
\[
(\pf M)^2=\det M.
\]
The analogue of \eqref{1-rank} is given by
\ba\label{2-rank}
(M^{-1})_{ji} \pf (M)
=
\int D\xi\,
\exp\!\left(
\frac12\xi^TM\xi
\right)\xi_i \xi_j ,
\ea
which leads to the Pfaffian counterpart of \eqref{matrix-det} -- the rank-two Pfaffian identity:
\ba\label{Pfaffian-rank-2}
\pf (M + X Y^T - YX^T) = \int D\xi\,
\exp\!\left(\frac12\xi^T (M + X Y^T - YX^T) \xi\right) \nonumber\\
=\int D\xi\,
\exp\!\left(\frac12\xi^T M\xi\right) (1 + \xi^T  X Y^T\xi )
\stackrel{\eqref{2-rank}}{=} \pf (M)\, (1 + Y^T M^{-1} X).
\ea

Also, it is not hard to derive the expansion of its Pfaffian with respect to the first row in complete analogy with the usual determinant expansion.
Thus we have
\[
\frac12\xi^T M\xi
=
\xi_1\sum_{j=2}^{2n}M_{1j}\xi_j
+
\frac12\sum_{i,j=2}^{2n}\xi_i M_{ij}\xi_j = \frac12{\xi'}^T M'\xi'
+
\xi_1 L,
\]
where \(
\xi'=(\xi_2,\ldots,\xi_{2n})^T,
\
M'=(M_{ij})_{i,j=2}^{2n},
\)
and
\(
L :=\sum_{j=2}^{2n}M_{1j}\xi_j.
\)

Since \((\xi_1L)^2=0\), we get
\[
\exp\!\left(\frac12\xi^T M\xi\right)
=
\exp\!\left(\frac12{\xi'}^T M'\xi'\right)
\bigl(1+\xi_1L\bigr).
\]
Only the term containing \(\xi_1\) contributes to the Berezin integral.
Therefore,
\[
\pf M
=
\sum_{j=2}^{2n}M_{1j}
\int d\xi_{2n}\cdots d\xi_2\,
\exp\!\left(\frac12{\xi'}^T M'\xi'\right)\xi_j .
\]

For a fixed \(j\), only the term containing all variables except
\(\xi_j\) contributes. Reordering the resulting Grassmann monomial into
the canonical order produces the sign
\(
(-1)^{2n-j}=(-1)^j.
\)
Hence
\[
\int d\xi_{2n}\cdots d\xi_2\,
\exp\!\left(\frac12{\xi'}^T M'\xi'\right)\xi_j
=
(-1)^j\pf(M_{\widehat{1}\widehat{j}}),
\]
where \(M_{\widehat{1}\widehat{j}}\) denotes the antisymmetric matrix
obtained by deleting the first and the \(j\)-th rows and columns of \(M\).
Finally,\footnote{The expansion with respect to an arbitrary row \(i\) is
\[
\pf(M)
=
\sum_{j\neq i}
(-1)^{\,i+j+\mathbf 1_{\{j>i\}}}
M_{ij}\,
\pf(M_{\widehat i\widehat j}),
\]
where the sign comes from first moving the \(i\)-th row and column to the first position.}
\ba\label{Pfaffian:expansion}
\pf M
=
\sum_{j=2}^{2n}
(-1)^j M_{1j}\,
\pf(M_{\widehat{1}\widehat{j}}).
\ea

\bibliographystyle{elsarticle-num}
\bibliography{pfafinal_r1}
\end{document}